\documentclass[letterpaper]{jpconf}
\usepackage{graphicx}
\usepackage{floatflt}
\usepackage{wrapfig}

\begin{document}
\title{Measurement of the longitudinal structure function of the proton, $F_L$}

\author{Tim Namsoo - on behalf of the ZEUS and H1 collaborations}

\address{DESY}

\begin{abstract}
The ZEUS and H1 collaborations have made the first direct measurements of the longitudinal proton structure function, $F_L$, which is strongly correlated to the gluon density in the proton. The ZEUS collaboration have also extracted the structure function, $F_2$, which, for the first time, has been done so at high $y$ without any assumptions about $F_L$.
\end{abstract}

\section{Introduction}
The inclusive $e^\pm p$ deep inelastic scattering (DIS) cross section, $\sigma^{e^\pm p}$, can be expressed in terms of the two structure functions, $F_2$ and $F_L$, so long as the virtuality of the exchanged boson, $Q^2$, is small enough such that weak interactions can be ignored. Symbolically,
\begin{equation}
\frac{d^2\sigma^{e^\pm p}}{dxdQ^2} = \frac{2\pi\alpha^2Y_+}{xQ^4}
\left[F_2(x,Q^2) - \frac{y^2}{Y_+} F_L(x,Q^2)\right]
= \frac{2\pi\alpha^2Y_+}{xQ^4} \; \tilde{\sigma} (x,Q^2,y),
\label{eq:disnc-xsec}
\end{equation}
where $\alpha$ is the fine structure constant, $x$ is the Bjorken scaling variable, $y$ is the inelasticity and $Y_+=1+(1-y)^2$. The quantity $\tilde{\sigma}$ is known as the reduced cross section. 

The structure function $F_2$ is proportional to the total virtual-photon/proton scattering cross section, whereas $F_L$ is only sensitive to its longitudinally polarised virtual-photon component. This is only non-zero due to gluon exchange within the proton, hence $F_L$ is strongly correlated to the gluon density in the proton.  The ratio $R=F_L/(F_2-F_L)$, is the ratio of the longitudinally- to transversely-polarised virtual-photon/proton scattering cross sections. 

The reduced $ep$ DIS cross section and related $F_2$~\cite{Chekanov:2001qu,Adloff:1997mf} measurements at HERA provide the strongest constraints on the proton parton distribution functions (PDFs) at low $x$. Within the DGLAP QCD formalism, $F_2$ directly relates to the valence quark and $q\bar{q}$ sea distributions, while sensitivity to the gluon distribution, $g(x,Q^2)$, is achieved through the scaling violations in $F_2$ at low $x$. However, fits of $g(x,Q^2)$ (or equivalently indirect extractions of $F_L$~\cite{Adloff:1996yz,Adloff:2003uh}) derived from the scaling violations depend on the precise formalism adopted. In the same vein, all previously published values of $F_2$ have required assumptions to be made about $F_L$ at low $x$.  

The ZEUS and H1 collaborations have, for the first time, directly measured $F_L$, thus providing an important check of the QCD formalisms. Moreover, by extracting $F_2$ and $F_L$ simultaneously, ZEUS have made the first measurement of $F_2$ that did not require any QCD assumptions at low $x$/high $y$.

\section{The experimental method}
The direct $F_L$ extractions required both collaborations to measure the reduced cross sections at fixed $(x,Q^2)$ spanning a range of $y$. In $ep$ scattering, the centre-of-mass energy is given by $\sqrt{s}=\sqrt{Q^2/xy}$, therefore these measurements required data with multiple $\sqrt{s}$ values. The structure functions $F_2$ and $F_L$ were then evaluated in each bin of $x$ and $Q^2$ using a Rosenbluth plot~\cite{Rosenbluth:1950yq}, in which a straight line is fit to $\tilde{\sigma}(y^2/Y_{+})$. This is motivated by Eq.~\ref{eq:disnc-xsec}, which implies that $F_2(x,Q^2) = \tilde{\sigma}(x,Q^2,y=0)$, the intercept, and $F_L(x,Q^2) = -{\rm \partial}\tilde{\sigma}(x,Q^2,y)/{\rm \partial}(y^2/Y_{+})$, the slope.

The precision of this procedure depends on the range of $y^2/Y_{+}$ spanned by the data. This was maximised by collecting data at the nominal HERA energy, $\sqrt{s}=318$~GeV, and at $\sqrt{s}=225$~GeV, which was the lowest possible energy at which adequate instantaneous luminosity could be delivered. Data were also collected at $\sqrt{s}=251$~GeV. The change in the $\sqrt{s}$ was achieved by varying the proton beam energy, $E_{\rm p-beam}$, keeping the electron beam energy constant, $E_{\rm e-beam}=27.5$~GeV. Data were collected in 2006-07 with $E_{\rm p-beam}=920$, $575$ and $460$~GeV, referred to respectively as the HER, MER and LER samples, standing for high-, medium- and low-energy-running. 

\begin{wrapfigure}{r}{9.3cm}
\begin{center}
\vspace{-1.2cm}
\includegraphics[angle=0,scale=0.49]{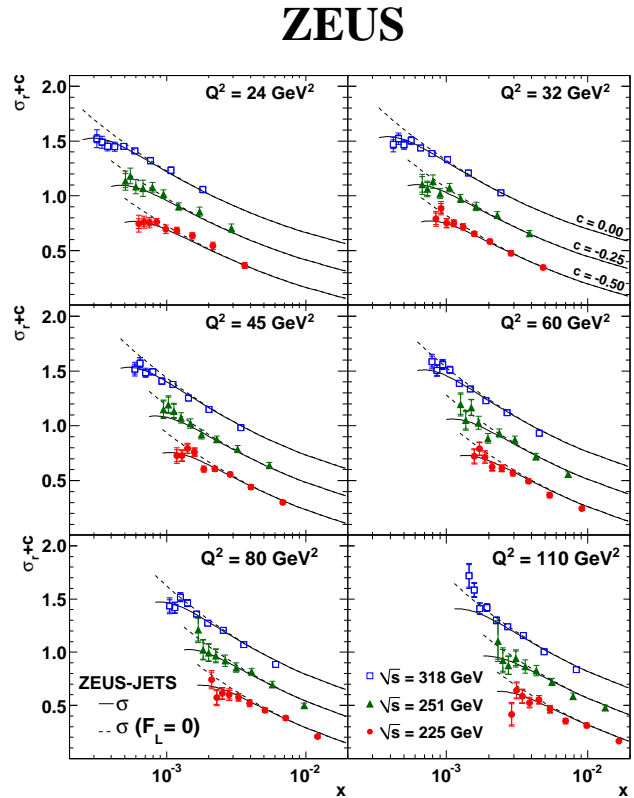}
\vspace{-1.3cm}
\caption{The HER, MER and LER reduced cross sections, shifted by $c_i$ (see top right) for clarity.  The inner (outer) error bars on the data represent the statistical uncertainties (statistical and systematic uncertainties added in quadrature).  
  \label{fig:ZEUSredXsec}}
\vspace{-1.3cm}
\end{center}
\end{wrapfigure}
Both experiments reconstructed the event kinematics based purely on the measured scattered electron. At high $y$, the scattered electron tends to be low in energy and poorly separated from the hadronic final state.  In this case, photoproduction is a significant background due to the misidentification of hadrons as electrons.  Both collaborations remove this on a statistical basis, however, H1 derived the background spectrum using wrongly charged electron candidates, while ZEUS used a Monte Carlo simulation.

\section{The results}
The H1 measurement was conducted in two parts, referred to as the mid- and high-$Q^2$ analyses\footnote{Additional H1 analyses have been made preliminary since LLWI09: a low-$x$ analysis, extending the kinematic reach down to $x=5.9\times10^{-5}$ and a diffractive $F_L$ analysis.  These analyses wont be discussed here.}.  The distinction is driven by the specific sub-components of the H1 calorimeter used in each analysis. The published mid-$Q^2$ analysis~\cite{Aaron:2008tx} spans the kinematic region, $12<Q^2<90~{\rm GeV^2/c^4}$ and $2.4\times10^{-4}<x<3.6\times10^{-3}$.  The preliminary high-$Q^2$ analysis spans the kinematic region, $35<Q^2<800~{\rm GeV^2/c^4}$ and $2.8\times10^{-4}<x<3.53\times10^{-2}$.  

The ZEUS $F_L$ measurement~\cite{ZEUSFL} spans the kinematic region, $24<Q^2<110~{\rm GeV^2/c^4}$ and $5\times10^{-4}<x<7\times10^{-3}$.  In addition to $F_L$, ZEUS also published the HER, MER and LER reduced cross sections, shown in Fig.~\ref{fig:ZEUSredXsec} at six $Q^2$ values as a function of $x$, and the first $F_2$ values extracted at low $x$/high $y$ without implicit $F_L$ assumptions.  Also shown in Fig.~\ref{fig:ZEUSredXsec} are predictions based on the NLO ZEUS-JETS proton PDF set~\cite{Chekanov:2005nn}, with both the expected $F_L$ contribution and with it set to zero. This indicates how $F_L$ is expected to suppress the cross sections at low $x$, the mechanism by which the experiments gain sensitivity to $F_L$. The data turn over at low $x$ as expected.
%
\begin{figure}
\begin{center}
\vspace{-0.5cm}
\hspace{-1.cm}
\includegraphics[angle=0,scale=0.6]{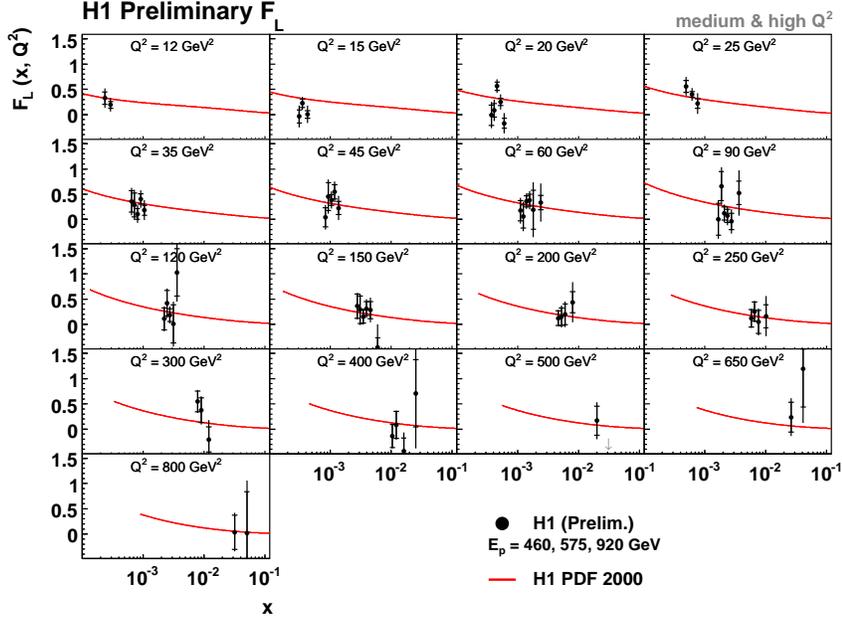}
\vspace{-1.0cm}
\caption{$F_L$ as a function of $x$ and $Q^2$. The $F_L$ prediction based on the NLO H1 PDF 2000 set is also shown. Other details as in the caption to Fig.~\ref{fig:ZEUSredXsec}.
  \label{fig:H1FLxQ2}}
\vspace{-0.8cm}
\end{center}
\end{figure}
%
\begin{wrapfigure}{r}{9.3cm}
\begin{center}
\vspace{-0.5cm}
\includegraphics[angle=0,scale=0.45]{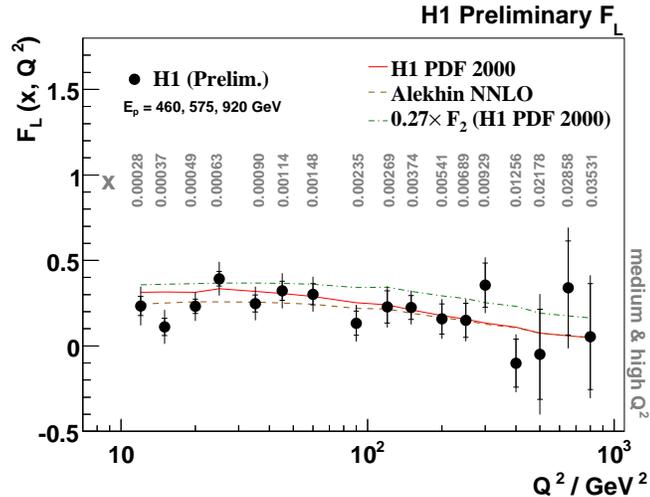}
\vspace{-0.5cm}
\caption{$F_L$ as a function of $Q_2$ at the given values of $x$. Shown too are $F_L$ predictions based on the NLO H1 PDF 2000 and NNLO Alekhin~\cite{Alekhin:2005gq} PDF sets and the H1 PDF 2000 $F_2$ prediction scaled by 0.27.  Other details as in the caption to Fig.~\ref{fig:ZEUSredXsec}.
  \label{fig:H1FLxQ2av}}
\vspace{-0.8cm}
\end{center}
\end{wrapfigure}

Once the reduced cross sections are measured, $F_L$ can be extracted. Shown in Fig.~\ref{fig:H1FLxQ2} are results from the combined mid- and high-$Q^2$ H1 analyses, given as a function of $x$ at fixed $Q^2$ points. Also shown is the $F_L$ prediction based on the NLO H1 PDF 2000 set~\cite{Adloff:2003uh}.  Generally, the agreement between theory and data is good. In most cases the $F_L$ points are positive. 

More precise $F_L$ values can be obtained by averaging these data.  The results are shown in Fig.~\ref{fig:H1FLxQ2av} as a function of $x$ and $Q^2$. Clearly, $F_L$ is observed to be non-zero over much of the $(x,Q^2)$ range probed.  Also included in the plot are $F_L$ predictions based on an NLO and NNLO PDF set, and an NLO $F_2$ prediction, scaled by 0.27.  Each of the curves are consistent with the data. 

\begin{wrapfigure}{r}{9.3cm}
\begin{center}
\vspace{-1.2cm}
\hspace{-0.8cm}
\includegraphics*[angle=0,scale=0.47]{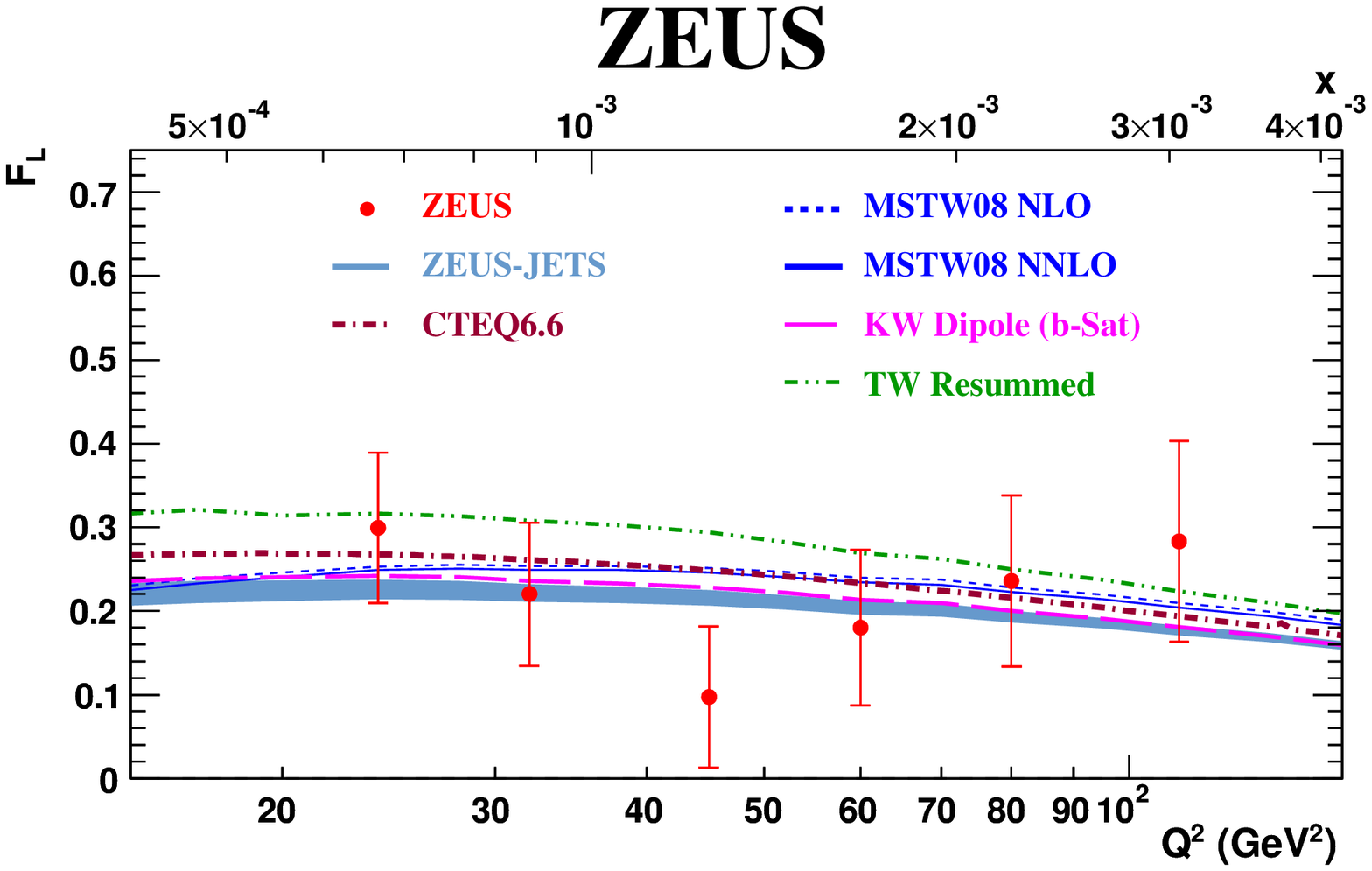}

\vspace{-0.3cm}
\hspace{-0.8cm}
\includegraphics*[angle=0,scale=0.47]{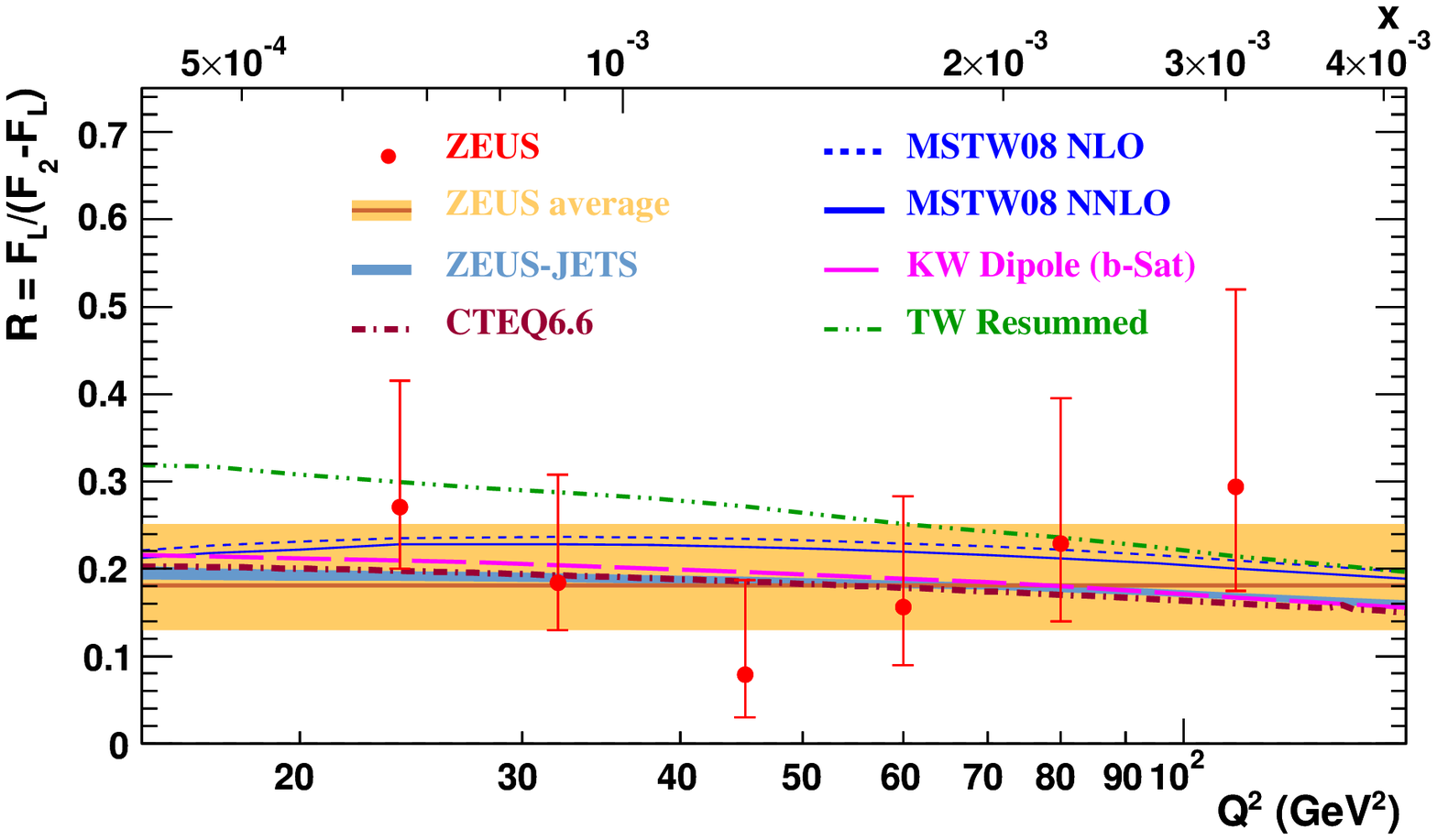}

\vspace{-7.1cm}
\hspace{7.5cm}
(a)

\vspace{4.9cm}
~(b)
\vspace{2.2cm}
\hspace{-7.5cm}

\vspace{-1.4cm}
\caption{
$F_L$ (a) and $R$ (b) as functions of $Q_2$ at the given values of $x$.  The error bars represent the combined statistical and systematic uncertainties.  
Shown too are predictions based on the ZEUS-JETS and CTEQ6.6~\cite{Nadolsky:2008zw} NLO and MSTW08~\cite{Martin:2009iq} NLO and NNLO PDF fits as well as a resummed fit from Thorne and White~\cite{White:2006yh} and a fit based on a colour glass condensate dipole model from Kowalski and Watt~\cite{Watt:2007nr}.
  \label{fig:ZEUSFLxQ2}}
\vspace{-1.5cm}
\end{center}
\end{wrapfigure}

The equivalent data from ZEUS are shown in Fig.~\ref{fig:ZEUSFLxQ2}a, albeit in a restricted kinematic region. Again, $F_L$ is observed to be positive. The extraction method used by ZEUS, while fully accounting for the correlations in the uncertainties, blurs the distinction between the statistical and systematic components.  The ZEUS results are consistent with those of H1. The data have been compared to theoretical predictions (see figure caption), the majority of which follow the DGLAP formalism at either NLO, NNLO or using an impact dependent dipole saturation model approach. The resummed prediction is based on the BFKL formalism.  The predictions are all consistent with the data.  

Shown in Fig.~\ref{fig:ZEUSFLxQ2}b is the ZEUS measurement of the ratio $R=F_L/(F_2-F_L)$ as a function of $x$ and $Q^2$, as well as a band indicating the 68\% probability interval for the overall value of $R$ in the kinematic region studied, $R=0.18^{+0.07}_{-0.05}$.   The data are compared to predictions based on the same set of PDFs used in Fig.~\ref{fig:ZEUSFLxQ2}a.  Again, all of the predictions are consistent with the data.

\section{Summary}
The ZEUS and H1 collaborations have published measurements of the structure function $F_L$.  The combined H1 analyses span a wide kinematic space.  The results from both collaborations are consistent in the region where they overlap.  The ZEUS collaboration have also published the reduced cross sections at $\sqrt{s}=\{318,251,225\}$~GeV and the the first $F_2$ values extracted without implicit $F_L$ assumptions.  The measurements are consistent with predictions based on different QCD formalisms.
\\
\\
\\
\\
\\

%
%

\end{document}